\documentclass{ws-procs11x85}

\usepackage{balance}
\usepackage{amsmath}
\usepackage{amssymb}
\usepackage{multirow}
\usepackage{dcolumn}
\newcolumntype{d}{D{.}{.}{-1}}
\newcolumntype{P}{D{,}{\,\pm\,}{-1}}

\usepackage{lp2003-yabsley-macro}

\makeindex
\begin{document}

\title{EXPERIMENTAL LIMITS ON NEW PHYSICS FROM CHARM DECAY}

\author{B. D. YABSLEY}

\address{Virginia Polytechnic Institute and State University,
	Blacksburg, VA 24061, USA				\\
	E-mail: yabsley@bmail.kek.jp				\\
        Web:    http://belle.kek.jp/$\,\tilde{~}$yabsley/}


\twocolumn[\maketitle\abstract{
Recent measurements in the charm sector are reviewed, concentrating on 
results which are sensitive to New Physics effects. 
The scope of the presentation includes $\dz-\dzbar$ mixing searches,
a CPT / Lorentz invariance study, and a range of searches for rare and
forbidden decays. Results from the BaBar, Belle, CDF, CLEO, and FOCUS
collaborations are presented, including an important first observation.}]


\baselineskip=13.07pt
\section{Introduction}
\label{sec:intro}

Presentations on ``New Physics'' can produce a feeling of anti-climax,
for surely if there were any New Physics signals to report, the news
would have leaked out.  One does not expect to hear anything new.
This talk does contain at least one first observation, however, and I'll
try to maintain some suspense by not mentioning in advance what it is.

\subsection{What I'm \underline{Not} Talking About}
\label{subsec:intro-not}

Today we are at a particular disadvantage because the year's biggest
charm news is outside the scope of the talk.
BaBar's discovery\cite{dsj-babar} of a narrow resonance decaying to
$D_s \pi^0$ came as a complete surprise---apart from the familiar 
$D_s^\ast$, no mesons decaying to this final state were foreseen---and
led to a flurry of speculation that the new $D_{sJ}(2317)$
might be an exotic meson. The honors were shared among the $B$-factories
in an amusing way: CLEO announced the discovery\cite{dsj-cleo}
of a second state, the $D_{sJ}(2457)$, decaying to $D_s^\ast \pi^0$;
and Belle, as well as confirming these results,\cite{dsj-belle-ccbar}
made the first observation of both states
in $B$ meson decays.\cite{dsj-belle-bmes}
The decay modes and widths of the new $D_{sJ}$ are consistent\footnote{
	In the case of the $D_{sJ}(2457)$, the Belle $D_s \gamma$
	results\cite{dsj-belle-bmes,dsj-belle-ccbar} rule out $J=0,\,2$,
	and are consistent with $J=1$.
	Note that an important ``exotic'' hypothesis---that the
	$D_{sJ}(2317)$ is wholly or partly a $D K$ bound state, and the
	$D_{sJ}(2457)$ likewise a $D^\ast K$ state---is also consistent 
	with the $0^+$ and $1^+$ assignments.}
with these states
being the $J^P = 0^+$ ($D_{sJ}(2317)$) and $J^P = 1^+$ ($D_{sJ}(2457)$)
members of the $c\bar{s}$ system (with $L=1$, and ``light quark angular
momentum'' $j_q=\frac{1}{2}$), but their masses are a complete mystery.
We thought we understood the $c\bar{q}$ mesons \ldots but it appears that 
we don't. This topic will be covered further in Jussara de Miranda's talk
on Standard Model charm studies.\cite{lp03-miranda}

The largest discrepancy between theory and experiment in the charm sector
is also off-limits, as no-one imagines that new physics is responsible.  
But we still don't understand why the measured
cross-section\cite{cccc-belle-prl,cccc-belle-conf}
for $e^+e^- \to \psi\,\eta_c$ is an order of magnitude larger than the
NRQCD prediction; the ingenious suggestion\cite{cccc-bbl} of
$e^+e^- \to \gamma^\ast \gamma^\ast \to \psi\,\psi$ contamination has
now been ruled out.\cite{cccc-belle-comment}
The $\psi\,\ccbar$ fraction in $e^+e^- \to \psi\,X$ is likewise far 
``too large'', almost saturating $\psi$ production.
We had thought that we understood \ccbar\ production at this energy
\ldots but it's pretty clear that we don't.
Tomasz Skwarnicki will have something to say about this, and other
developments in charmonia, in the next presentation.\cite{lp03-skwarnicki}

\subsection{What I \underline{Am} Talking About}
\label{subsec:intro-am}

After discussing the problems of obtaining clean new physics
signatures in the charm sector (Sec.~\ref{sec:signatures}),
the largest part of the talk treats $\dz-\dzbar$ mixing
searches (Sec.~\ref{sec:mixing}); there are important new results
on both \ycp\ (Sec~\ref{subsec:mixing-ycp})
and $\dz \to K^+ \pi^-$ (Sec.~\ref{subsec:mixing-kpi}).
The first CPT and Lorentz invariance violation study in charm has
recently been published, and we discuss it briefly (Sec.~\ref{sec:cpt}). 
The remainder of the talk is given over to searches for rare and 
forbidden decays (Sec.~\ref{sec:rare}).
There are new results from BaBar, Belle, CDF, CLEO, and FOCUS---including,
as I say, a first observation---but let's not get ahead of ourselves.


\section{Finding Clear New Physics Signatures}
\label{sec:signatures}

The difficulty with finding a clear signature of New Physics in the charm
sector is this: it can be hard to know what the Standard Model (SM) 
prediction is. One way of thinking about the problem is to consider the
masses of the quarks.
\begin{itemize}
  \item	The up and down quark masses are both small compared to the
  	hadronic mass scale: $m_u < m_d \ll \lambda_{QCD}$.\footnote{
		We set aside the related fact that $m_u$ and $m_d$ 
		are elements of the theory,
		rather than straightforward ``observations''.}
	Isospin is therefore a rather good symmetry, and (including
	now the strange quark) $SU(3)$ of flavor, while broken, is 
	useful.
  \item	At the other extreme, the beauty quark has $m_b > \lambda_{QCD}$,
  	and can be considered as a high-energy physics ``particle'':
	a ``billiard ball with quantum numbers attached''. One can think
	in terms of the Feynman diagrams, in $b$-sector
	processes, and not be too seriously misled.
  \item	The charm quark lies between the two extremes:
	$m_c \gtrsim \lambda_{QCD}$, neither light nor truly heavy.
\end{itemize}
The awkwardness of the charm mass thus puts limits on both symmetry- and 
quark-based thinking as guides to charm physics. If light hadron work
is like swimming in the ocean, and $b$-physics is like flying through
the air, then in charm studies one is wading knee-deep through the 
brown muck.

One should really think in terms of hadrons, not quarks,
in charm. So-called ``long-distance'' contributions
are important in many processes: quark loops are typically
suppressed, so that hadronic processes take a leading role.
These are usually difficult to calculate, especially as the charm
mass lies in the resonance region.
So if some parameter is supposed to be small, but observed to be large,
one should be cautious before claiming new physics:
perhaps the Standard Model contribution has just been miscalculated.
As discussed in Sec.~\ref{subsec:intro-not}, there have already been two major
surprises in the last two years.


\section{$\dz-\dzbar$ Mixing}
\label{sec:mixing}

\subsection{Mixing in the Standard Model and Beyond}
\label{subsec:mixing-sm}

\begin{figure}
  \center
  \psfig{figure=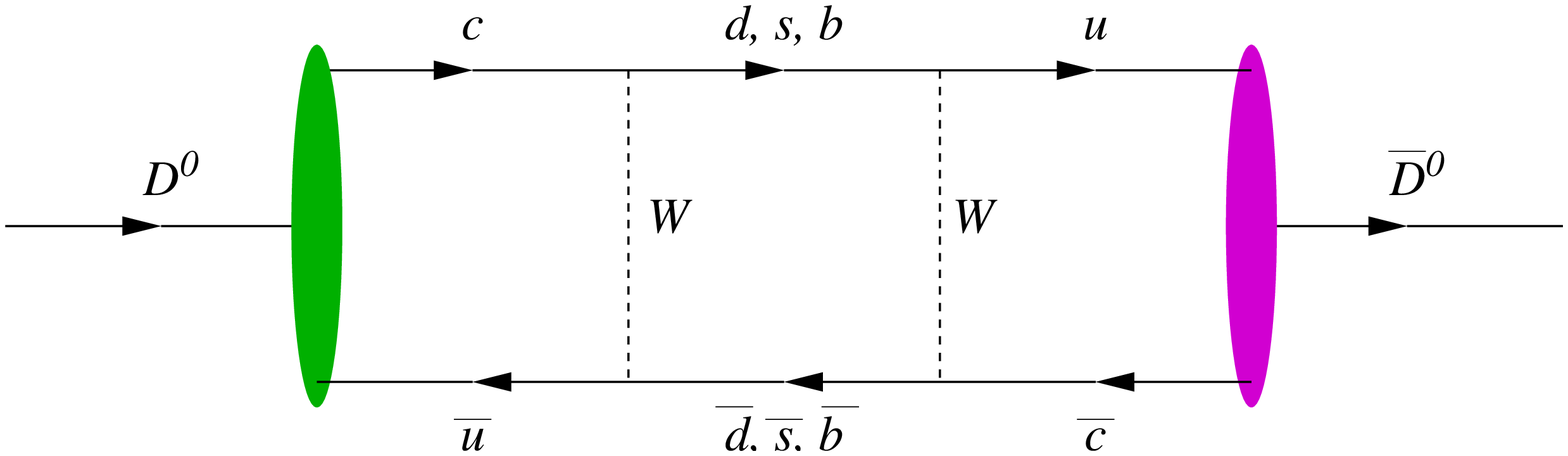,width=8.0truecm}
  \caption{Box diagram for $\dz-\dzbar$ mixing in the SM.}
  \label{fig:dmixing-box-3gen}
\end{figure}

\begin{figure}
  \center
  \psfig{figure=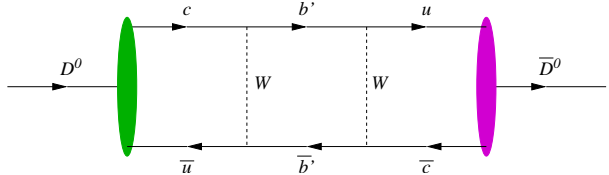,width=8.0truecm}
  \caption{Box diagram for $\dz-\dzbar$ mixing in a New Physics model with an extra down-type quark.}
  \label{fig:dmixing-box-4gen}
\end{figure}

There are particular pitfalls in the interpretation of charm mixing
searches. The SM box diagrams for mixing
(e.g.\ Fig.~\ref{fig:dmixing-box-3gen}) are doubly Cabibbo-suppressed
and suffer from very efficient cancellations (the GIM mechanism):
the expected mixing rate due to such processes is negligible. 
Since most new physics scenarios introduce new particles that couple to the SM
fields, they induce new loop diagrams such as Fig.~\ref{fig:dmixing-box-4gen},
with no (or a lesser degree of) cancellations.
The result is an enhancement of the mass splitting of the $\dz-\dzbar$
eigenstates, $x \equiv \Delta M / \Gamma$; hence the common
statement that
``$\dz-\dzbar$ mixing with measurable $x$ is a signal of New Physics''.

\begin{figure}
  \center
  \psfig{figure=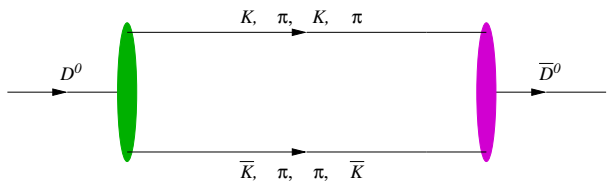,width=8.0truecm}
  \caption{Sample diagram for $\dz-\dzbar$ mixing due to common hadronic final states.}
  \label{fig:dmixing-box-kpi}
\end{figure}

As discussed in the previous section, however, hadronic processes cannot
be neglected. Final states common to \dz\ and \dzbar, such as
$K\overline{K}$, $\pi\pi$, $K\pi$ and $\overline{K}\pi$, couple the two
neutral $D$'s (Fig.~\ref{fig:dmixing-box-kpi}); such contributions cancel
in the $SU(3)_F$ limit, but to the extent that $SU(3)$ of flavor is broken,
they induce mixing.
One might suppose that only the lifetime splitting
parameter $y \equiv \Delta \Gamma / 2\Gamma$ would be affected,
as the intermediate states are real. The true situation is more complicated.
To the extent that quark-hadron duality holds, mixing can be
estimated using the Operator Product Expansion:
a recent study\cite{bigi} finds $x \sim y \sim O(10^{-3})$.
An alternative approach\cite{falk} relying directly on hadronic intermediate
states suggests that $y$ may be as large as $O(1\%)$. 
So as far as $x$ and $y$ are concerned, mixing provides
a clean new physics signal only if $x \gg y \sim 10^{-3}$.

\begin{figure*}
  \center
  \psfig{figure=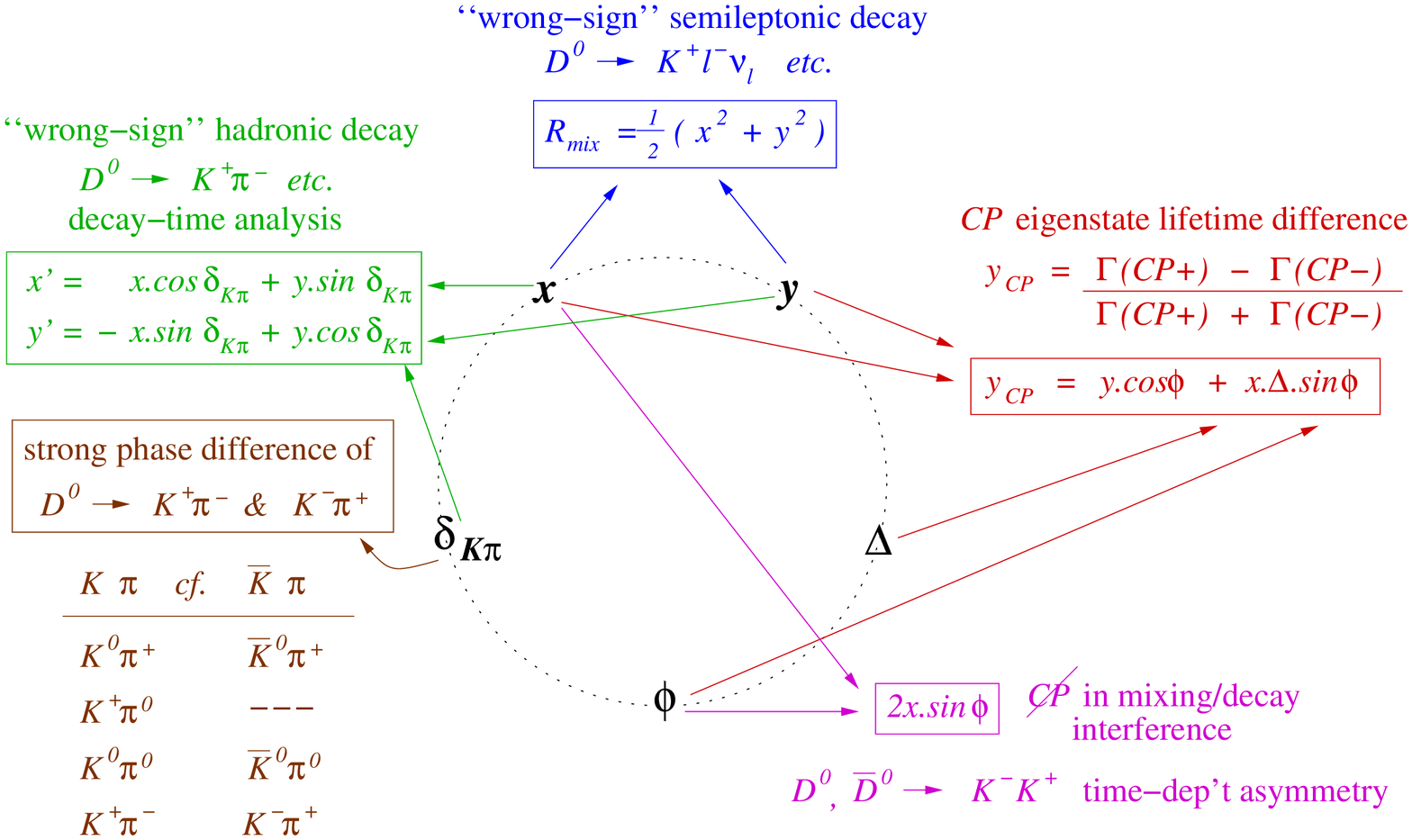,width=14.0truecm}
  \caption{Five parameters describing $\dz-\dzbar$ mixing, and their relationship with
  	quantities measured in experiment.}
  \label{fig:dmixing-parameters}
\end{figure*}

\subsection{Mixing Parameters \& Measurements}
\label{subsec:mixing-params}

The full parameter space for mixing is more rich than $(x,y)$:
some important parameters are shown in Fig.~\ref{fig:dmixing-parameters},
together with experimentally measurable quantities.
The parameters are poorly-known:
the strong phase difference $\delta_{K\pi}$ between $\dz \to K^+ \pi^-$
and $K^- \pi^+$ amplitudes is unconstrained,
as is the difference in particle
content of the eigenstates $|D_{1,2}\rangle =  p |\dz\rangle \pm q |\dzbar\rangle$,
$\Delta = (|p|^2 - |q|^2)/(|p|^2 + |q|^2)$.
But the news is not all bad.
This year has seen the first measurements
relevant to the CP violating phase
$\phi=\arg\left(q {\mathcal A}(\dzbar \rightarrow K^- K^+) /
		p {\mathcal A}(\dz    \rightarrow K^- K^+) \right)$
(Sec.~\ref{subsub:mixing-ycp-cpv}),
as well as a major new analysis of $\dz \to K^+ \pi^-$
(Sec.~\ref{subsec:mixing-kpi}).


\subsection{Mixing: Lifetime Difference, \ycp, and CP Violation}
\label{subsec:mixing-ycp}

The most popular measurement in recent years, however, has been \ycp.
Defined as the normalized lifetime difference of the \dz-\dzbar\ CP eigenstates,
it is typically measured using the non-eigenstate decay $\dz \to K^- \pi^+$
as a convenient reference:
\begin{align*}
  \ycp \equiv	\frac{\Gamma(CP+)-\Gamma(CP-)}{\Gamma(CP+)+\Gamma(CP-)}
	& \approx\frac	{\tau(\dz \to K^- \pi^+)}
			{\tau(\dz \to K^- K^+)} - 1	\\[1ex]
	& =	y \cos \phi + x \Delta \sin \phi,
\end{align*}
where the last relation holds for small values of the parameters.
In the CP-conserving limit $\phi = 0$ and $\Delta = 0$, so $\ycp = y$, 
as one would expect. In this limit \ycp\ is not a new-physics search parameter
(since new particles are expected to affect $x$; Sec.~\ref{subsec:mixing-sm})
but a tool for measuring the level of mixing due to the SM.

\subsubsection{\ycp: the FOCUS measurement (2000)}
\label{subsub:mixing-ycp-focus}

Three years ago, the FOCUS collaboration measured \ycp\ using a
relatively clean sample of 10,000 $\dz\to K^- K^+$ events,\footnote{
	Inclusion of charge-conjugate modes is implied throughout,
	unless the context makes clear that they are treated separately.}
and a $K^-\pi^+$ sample ten times that size.\cite{ycp-focus}
Both inclusive and \dstar-tagged decays were used, under FOCUS-standard
reconstruction, particle identification, and vertex detachment cuts;
the result was obtained from a binned maximum-likelihood (ML) fit
to the distributions of \emph{reduced proper time}
$t^\prime \equiv (l - N\sigma_l)/\beta\gamma c$,
where $l,\sigma_l$ are the \dz\ decay length and its error,
and $N$ the minimum required detachment of the production and decay vertices.


The result was surprisingly large:
$\ycp = (3.42 \pm 1.39 \pm 0.74)\%$, over $2\sigma$ away from zero. 
There was considerable excitement at the thought that charm mixing 
might be within our grasp, partly due to the usual association of mixing
with new physics. But as discussed above, the most natural interpretation 
of percent-level \ycp\ would be a large lifetime difference parameter $y$,
due to Standard Model effects. The new physics interpretation was possible,
but somewhat forced: mixing with $x \gg y$ and large CP violation,
$\sin\phi \sim O(1)$, $\Delta \sim O(1)$.

\subsubsection{\ycp: Belle and CLEO (2002)}
\label{subsub:mixing-ycp-belle-untag}

New measurements have followed in short order, using \dz\ produced in
$e^+e^- \to \ccbar$ interactions at the $B$-factories.
Belle\cite{ycp-belle-untag} used a sample somewhat larger and cleaner
than that of FOCUS, performing an unbinned maximum-likelihood (UML)
fit to inclusive $\dz \to K^-\pi^+$ and $K^+K^-$ decay time distributions,
and measuring $\ycp = \left( -0.5 \pm 1.0^{+0.7}_{-0.8} \right)\%$.
CLEO\cite{ycp-cleo} used a smaller sample (the $9.0\,\ifb$ CLEO II.V run),
required a \dstar-tag, and added $\dz\to \pi^+\pi^-$ to the usual modes;
they measured $\ycp = \left( -1.2 \pm 2.5 \pm 1.4 \right)\%$.
Both results are manifestly consistent with zero, and each other---and the
fact that both are negative has led many to discount the FOCUS 
result. But it's worth noting that the average \ycp\ from the three is 
positive, and $\sim 1\%$.

\subsubsection{\ycp: BaBar (2003)}
\label{subsub:mixing-ycp-babar}

\begin{figure}
  \center
  \psfig{figure=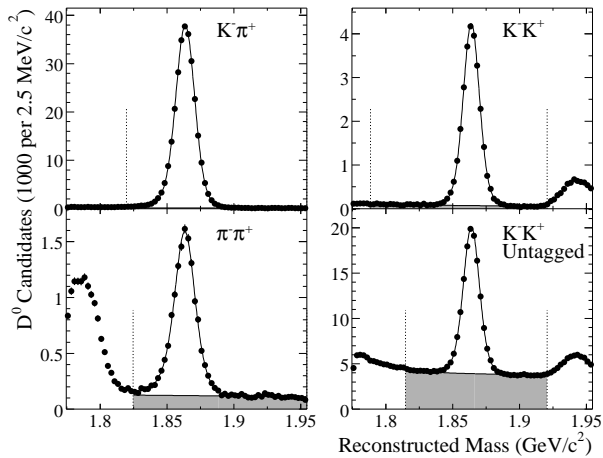,width=8.0truecm}
  \caption{BaBar \ycp\ analysis:\protect\cite{ycp-babar}
  	invariant mass distribution for data (points),
	projection of the fit (curve), and
	fitted background component (shaded)
	for the four samples.}
  \label{fig:ycp-babar-mass}
\end{figure}

This year BaBar has released a comprehensive \ycp\ measurement\cite{ycp-babar}
based on $91\,\ifb$ of data including both
\dstar-tagged $\dz \to K^-\pi^-,\, K^+K^-,\, \pi^+\pi^-$ events,
and inclusive $\dz \to K^+K^-$.
As usual in $e^+e^-$ analyses, backgrounds are suppressed by a
center-of-mass momentum cut and vertex quality cut on the \dz,
particle identification (PID) cuts on the daughter tracks,
and (for the \dstar-tagged samples) cuts on the \dstar-decay pion (the ``slow pion'').
The resulting \dz\ samples are shown in Fig.~\ref{fig:ycp-babar-mass}.
For each sample, the fitted mass distribution is used to determine the
event-by-event probability that a \dz\ candidate belongs to the
signal, as opposed to the background under the peak. 
This probability is then included in the likelihood function for each event
in an UML fit to the proper-time distribution.

\begin{figure}
  \center
  \psfig{figure=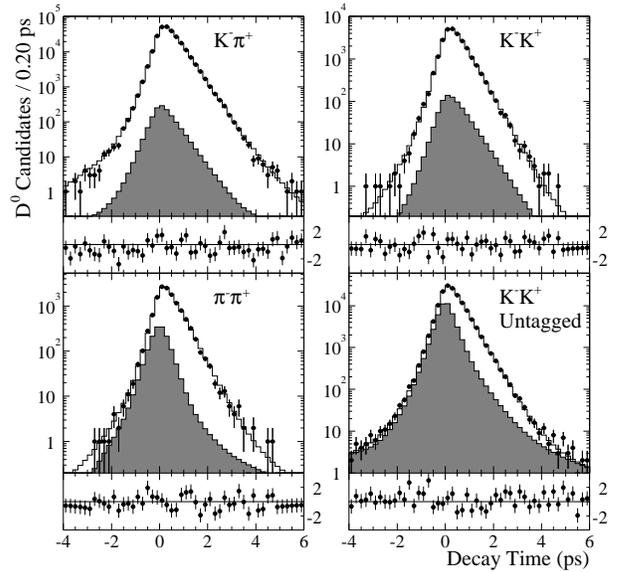,width=8.0truecm}
  \caption{BaBar \ycp\ analysis:\protect\cite{ycp-babar}
  	proper time distribution for data (points),
	UML fit projection (open histogram),
	and fitted background component (shaded)
	for the four samples. The points beneath each plot show the
	bin-by-bin differences between the data and the fit,
	divided by the statistical error.}
  \label{fig:ycp-babar-time}
\end{figure}

These distributions, and the results of the fits for each sample,
are shown in Fig.~\ref{fig:ycp-babar-time}. In each fit, the assumed
``underlying'' distributions of both signal and background are convolved
with resolution functions based on a sum of gaussians: 
most of these terms have widths of the form $S\sigma_t^i$,
where $\sigma_t^i$ is the event-by-event proper time error,
and $S$ is a scaling factor, meant to account for deficiencies in modelling
of the detector, etc.. This method is common to the earlier
Belle\cite{ycp-belle-untag} and CLEO\cite{ycp-cleo} analyses and reflects
a consensus on time-distribution fitting at the $B$-factories.

A blind analysis was performed to obtain the mixing parameter:
the weighted average over the four modes is
$\ycp = \left( 0.8 \pm 0.4^{+0.5}_{-0.4} \right) \%$,
the most precise measurement to date.

\subsubsection{\ycp: Belle, \dstar-tagged (2003)}
\label{subsub:mixing-ycp-belle-tag}

A new analysis from Belle, contributed to this symposium,\cite{ycp-belle-tag}
takes a different approach. The idea is to find a robust resolution function
which does not rely on the estimated proper-time error: this allows binned ML
fits to be used throughout, so that the goodness-of-fit can be explicitly
checked. \dstar-tagged $\dz\to K^- \pi^+$ and $K^+K^-$ events from the full
Belle dataset of $158\,\ifb$ are used, subject to standard reconstruction cuts,
and the requirement that the proper time be well-measured.

The $\dz$ lifetime, \ycp, and the parameters of the proper-time resolution
function are all determined in a single simultaneous binned ML fit to the
$K^- \pi^+$ and $K^+K^-$ samples.
The form of the resolution function is simple: a sum of five gaussians
with a common mean, fixed relative normalizations, and floating widths.
The gaussian widths for $K^+K^-$ are constrained to be the same as those
for $K^-\pi^+$, up to a single scale factor which is common to all terms.
This parameterization has been studied using Monte Carlo (MC) data,
and proves to be very stable: all values determined in a full decay-time fit
match those fitted to the true resolution function, within their relative errors
(Table~\ref{tab:ycp-belle-res}).

\begin{tablehere}
  \caption{Belle \ycp\ analysis:\protect\cite{ycp-belle-tag}
  	comparison of the parameters obtained in MC
	from a fit to the resolution function, using MC truth information (3rd column)
	and from the decay time fit, using reconstructed information only (4th column).
	The fractions of the five gaussian terms, fixed from the resolution function fit,
	are also shown (2nd column).}
  \label{tab:ycp-belle-res}
  $
  \renewcommand{\arraystretch}{1.2}
  \begin{array}
   {|cdPP|}
    \hline 
	&	& \multicolumn{2}{c|}{\text{fitted values (fs; except $\alpha$)}} \\
   \text{par.}	& \multicolumn{1}{c}{\text{fraction (\%)}}
				& \multicolumn{1}{c}{\text{ resolution fit}}
					& \multicolumn{1}{c|}{\text{lifetime fit}}\\
    \hline
    \sigma_1	& 26.1		&  95.1 ,  1.3		&  94.4 , 1.7	\\
    \sigma_2	& 50.4		& 177.0 ,  2.2		& 179.0 , 1.2	\\
    \sigma_3	& 19.8		& 328.7 ,  7.4		& 328.2 , 2.2	\\
    \sigma_4	& 3.1		& 675.7 , 24.9		& 664.4 , 8.5	\\
    \sigma_5	& 0.6		& 2199  , 95		& 2225  , 70	\\
    \hline
    X_0		& \multicolumn{1}{c}{\text{\footnotesize [common shift]}}
				& -1.51 , 0.22		& -0.95 , 0.54	\\ 
    \alpha	& \multicolumn{1}{c}{\text{\footnotesize [$K\pi \to KK$ scale]}}
				& 1.043 , 0.004		& 1.042 , 0.007	\\ 
    \hline
  \end{array}
  $
\end{tablehere}  

The proper time distribution for data is shown in Fig.~\ref{fig:ycp-belle-time},
together with the result of the binned ML fit; the confidence level is 94\%. 
(All fits in the analysis have an acceptable confidence level.)
The fit returns a \dz\ lifetime (from the $K^- \pi^+$ sample) of
$412.6 \pm 1.1\,\fs$, consistent with the world average,\cite{pdg}
and a mixing parameter $\ycp = (1.15 \pm 0.69 \pm 0.38)\%$;
the result is preliminary. 

\begin{figure}
  \center
  \psfig{figure=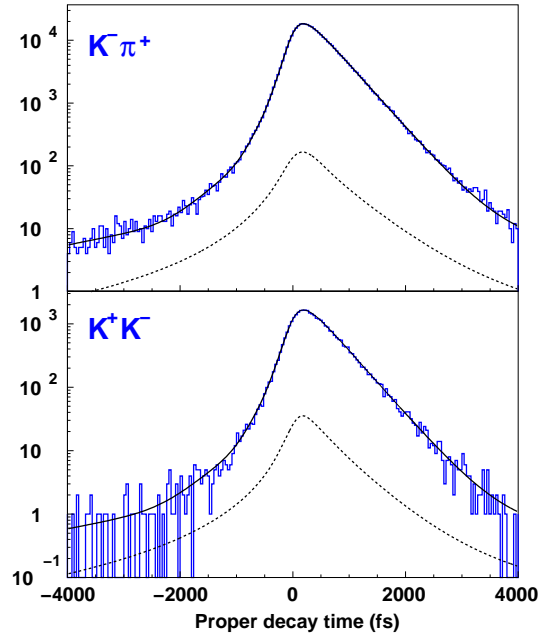,width=7.0truecm}
  \caption{Belle \ycp\ analysis:\protect\cite{ycp-belle-tag}
	proper time distribution for data (histogram),
	binned ML fit (solid curve), and
	background component (dashed),
	for the $K^- \pi^+$ (upper) and $K^+ K^-$ (lower) samples.}
  \label{fig:ycp-belle-time}
\end{figure}

\subsubsection{\ycp: summary of results}

For completeness, the \ycp\ results are summarized in Table~\ref{tab:ycp-summary}.
Note the dominance of the numbers from Belle (the tagged result,\cite{ycp-belle-tag}
as noted, is preliminary) and BaBar. The implications of these results were not
addressed at this point in the talk, but in the Discussion at the end. We preserve
this order here.

\begin{tablehere}
  \caption{Summary of \ycp\ results.}
  \label{tab:ycp-summary}
  \begin{center}
    \renewcommand{\arraystretch}{1.2}
    \begin{tabular}{|ll|}
	\hline
	Experiment			& \multicolumn{1}{c|}{\phantom{--}\ycp\ (\%)}	\\
	\hline
	E791\cite{ycp-e791}			& $\phantom{-}0.8 \pm 2.9 \pm 1.0$	\\
	FOCUS\cite{ycp-focus}			& $\phantom{-}3.4 \pm 1.4 \pm 0.7$	\\
	Belle, untagged\cite{ycp-belle-untag}	& $          -0.5 \pm 1.0 \pm 0.8$	\\
	CLEO\cite{ycp-cleo}			& $          -1.2 \pm 2.5 \pm 1.4$	\\
	BaBar\cite{ycp-babar}			& $\phantom{-}0.8 \pm 0.4^{+0.5}_{-0.4}$\\
	Belle, tagged\cite{ycp-belle-tag}	& $\phantom{-}1.2 \pm 0.7 \pm 0.4$	\\
	\hline
  \end{tabular}
  \end{center}
\end{tablehere}

\subsubsection{``$\ycp +\!+$'': \dz-\dzbar\ mixing and CPV}
\label{subsub:mixing-ycp-cpv}

In \ycp\ analyses based on \dstar-tagged samples, a CP-violation study
comes ``for free'': one can simply compare the lifetime of $K^+K^-$
events with different flavor $D$-tags, within the same analysis framework
used for \ycp. Belle\cite{ycp-belle-tag} defines a parameter
\begin{align*}
  A^{}_\Gamma	&\equiv	\frac{	\hat{\Gamma}(D     \to KK) -
  				\hat{\Gamma}(\dbar \to KK)}
			     {	\hat{\Gamma}(D     \to KK) +
				\hat{\Gamma}(\dbar \to KK)}	\\[.2truecm]
		&\approx -\Delta . y\cos\phi - x\sin\phi,
\end{align*}
where the notation $\hat{\Gamma}$ for ``effective lifetime'' recognises that 
an exponential is being fitted to distributions which may not be strictly
exponential.  
In the absence of CP violation in mixing, i.e. $\Delta = 0$, the
        asymmetry parameter $A^{}_\Gamma = -x\sin\phi$, measuring the CP
        violating phase $\phi = \arg(q\bar{A}/pA)$ due to the interference
        of decay and mixing.\footnote{
	The analysis is thus a close analogue of the $\bz/\bzbar \to \psi \ks$
	analysis in the $b$-sector, measuring $\sin 2\phi_1\, [\equiv \sin 2\beta]$.}
The ``$\Delta Y$'' parameter of BaBar\cite{ycp-babar} is similar, differing
by a factor of $(1+\ycp)$. Most systematic errors cancel due to the use of a
common final state.
The experiments find the following values,
\begin{align*}
  \Delta Y	& =	(-0.8 \pm 0.6 \pm 0.2)\%\;\; \text{(BaBar)}	\\
  A^{}_\Gamma	& =	(-0.2 \pm 0.6 \pm 0.3)\%\;\; \text{(Belle prelim.)},
\end{align*}
consistent with zero; the measurements are statistically dominated and will 
continue to improve for the life of the $B$-factories.

Unlike mixing in general, CP violation associated with mixing is a robust
new physics signal:\cite{bigi-sanda}
all charm mixing phenomena in the SM are dominated by the first two generations,
so CP violation must be small. 
Even for $x \sim y = O(1\%)$, we expect $A^{}_\Gamma \lesssim 10^{-4}$.
So any significant non-zero measurement by this technique would be evidence
of new physics contributing to mixing.
One can imagine a scenario where \emph{both} Standard Model and new physics
processes lead to percent-level values of the mixing parameters
($y$ and $x$ respectively), and the new physics contribution leads to 30\%-level
CP violation:
evidence for both SM mixing (via \ycp) and non-SM processes (via $A_\Gamma$)
would emerge by the end of the $B$-factory era.


\subsection{Mixing: $\dz \to K^+ \pi^-$}
\label{subsec:mixing-kpi}

Another new BaBar analysis,\cite{kpi-babar} of $\dz \to K^+ \pi^-$ decays, 
has brought hadronic mixing analyses back into prominence.
Pioneered by CLEO,\cite{kpi-cleo} the sophisticated analysis method is sensitive
to both mass- ($x$) and lifetime-splitting ($y$) of the neutral $D$ eigenstates,
and has been considered the technique of choice for $e^+e^-$ machines. 
The BaBar analysis has been gestating for some time---there were preliminary
presentations (without final fit results) two years ago---and the related 
analysis at Belle is still underway, with only an intermediate result
(the ``wrong-sign rate'' for $K\pi$) in the public domain.\cite{kpi-belle}

\subsubsection{$\dz \to K^+ \pi^-$: the analysis method}
\label{subsub:mixing-kpi-method}

``Wrong-sign'' hadronic decays such as $\dz \to K^+ \pi^-$ occur via two paths:
mixing $\dz \to \dzbar$ followed by Cabibbo-favoured decay $\dzbar \to K^+ \pi^-$, 
and directly by doubly-Cabibbo-suppressed (DCS) decay $\dz \to K^+ \pi^-$. 
The DCS decay thus forms a background to the mixing signal,
and the two must be separated by reconstructing the decay time of the \dz\
and exploiting the different time distributions: $e^{-t}$ for DCS decay
and $t^2 e^{-t}$ for mixing, where the proper time $t$ is in units of the \dz\ lifetime.
The interference term between the DCS and mixing paths, which goes as $t e^{-t}$,
cannot be neglected:\cite{kpi-blaylock} in fact it provides most of the 
mixing sensitivity, since the DCS rate is much larger than the mixing rate
and the interference term is thus intermediate in size.
A complication of the method is that this term is proportional,
not to the lifetime difference parameter $y$,
but to the quantity $y^\prime \equiv y\cos\delta_{K\pi} - x\sin\delta_{K\pi}$
which has been ``rotated'' by the strong phase difference $\delta_{K\pi}$
between the $\dz \to K^+ \pi^-$ and $K^- \pi^+$ decays.

The subtle difference in time distributions ($e^{-t}$, $t e^{-t}$, $t^2 e^{-t}$)
means that the time structure of background events must be well-understood
to avoid faking a mixing signal. This is important as background levels are 
relatively high. The method used by CLEO,\cite{kpi-cleo} which has been followed
by both BaBar\cite{kpi-babar} and Belle,\cite{kpi-belle} is to
(1) tag the initial $D$ flavor by reconstructing $\dstarp \to \dz \pi^+$;
(2) categorize the backgrounds according to their proper-time distribution; 
(3) measure their relative levels by fitting the data distribution in $(M,Q)$
where $M$ is the $K\pi$ mass and $Q = M(K^+\pi^-\pi^+) - M(K^+\pi^-) - m_\pi$ is
the energy release in \dstarp\ decay;\footnote{BaBar uses 
	$\delta m \equiv M(K^+\pi^-\pi^+) - M(K^+\pi^-)$, which differs from Q
	by a constant.} 
and
(4) fix the background levels in the fit to proper time.
This is difficult, but manageable, and CLEO reported limits at the 95\% confidence level
of $\frac{1}{2} x^{\prime 2} < 0.041\%$ and $-5.8\% < y^\prime < 1.0\%$ 
based on this method.

\subsubsection{$\dz \to K^+ \pi^-$: the BaBar analysis}
\label{subsub:mixing-kpi-babar}

The BaBar analysis follows the CLEO model closely,
but with a larger data sample ($57.1\,\ifb$),
lower backgrounds (due to BaBar's superior PID), and
a data selection and fitting procedure finalized while remaining ``blind''
to the mixing results.
The $K^+\pi^-$ distributions are shown in Fig.~\ref{fig:kpi-babar-mass}:
background events are divided into combinatorial, true-\dz-plus-unassociated-pion,
and double-misidentification categories.
This last type, where $K^-\pi^+$ is misidentified as $\pi^- K^+$, is small but significant:
see Fig.~\ref{fig:kpi-babar-mass}(b,d). Such events are retained to avoid any distortion
of the other backgrounds due to targeted rejection cuts.
The fit used to establish the background levels describes the data well.

\begin{figure}
  \center
  \psfig{figure=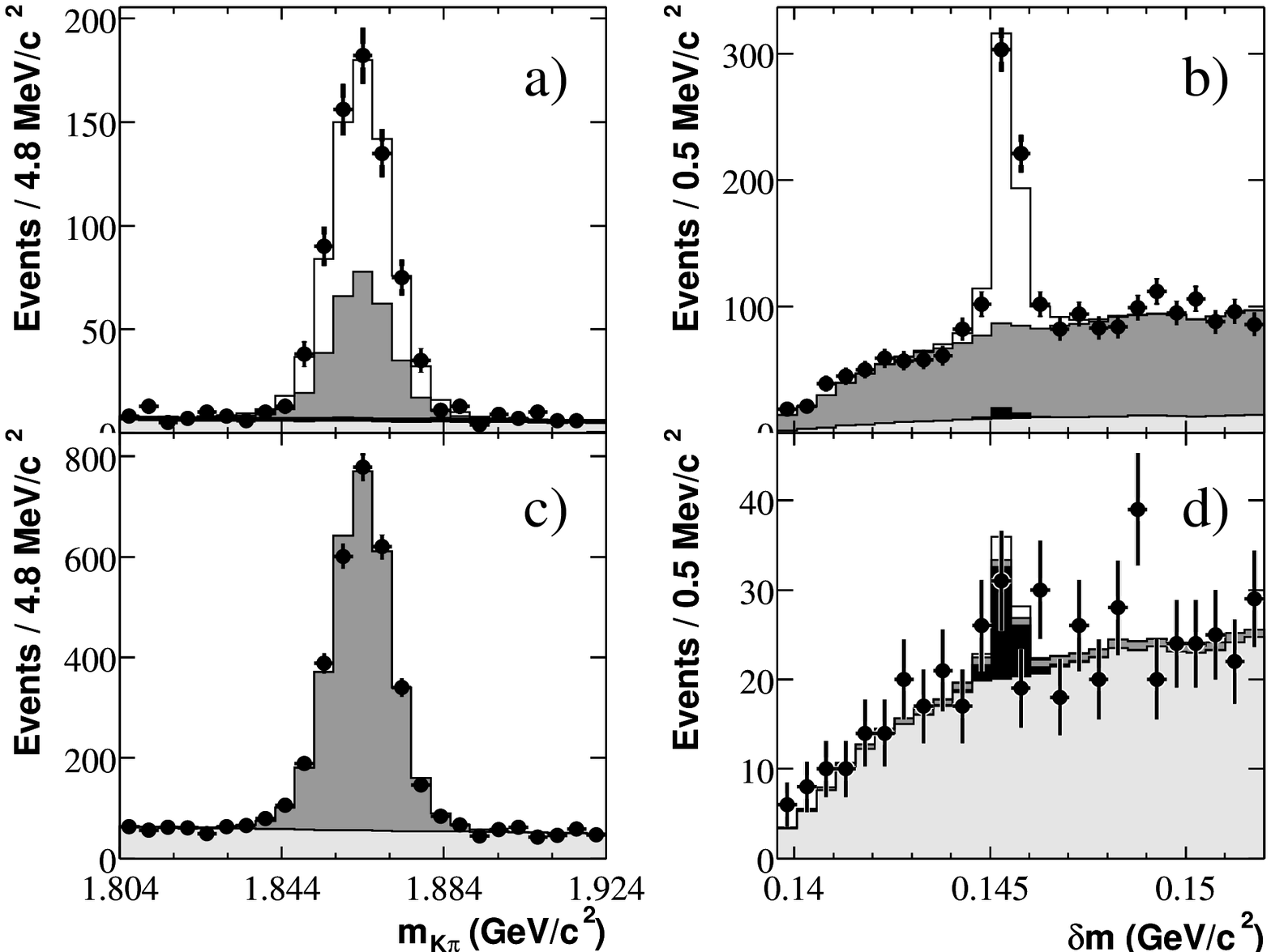,width=8.0truecm}
  \caption{BaBar $\dz\to K^+ \pi^-$ analysis:\protect\cite{kpi-babar}
	(a,c) $K^+\pi^-$ mass and (b,d) $\dstarp-\dz$ mass-difference distributions,
	for (a,b) signal- and (c,d) background-dominated regions.
	Shown are the data (points), the projection of the fit (open histogram),
	and the fitted combinatorial (light), unassociated pion (dark),
	and double-misidentification background (black).}
  \label{fig:kpi-babar-mass}
\end{figure}

The mixing parameters are then determined using unbinned, extended maximum-likelihood
fits to the $\dz \to K^+ \pi^-$ (``wrong sign'') and $\dz \to K^- \pi^+$ (``right sign'')
data. The likelihood terms for the signal and various background time distributions
are formed from underlying distributions (exponentials or delta functions) convolved with
event-dependent resolution functions similar to those of the \ycp\ analysis
(Sec.~\ref{subsub:mixing-ycp-babar}); the event-by-event signal- and background-fractions
are determined from the $(m_{K\pi},\delta m)$ fit.
The results are shown in Fig.~\ref{fig:kpi-babar-time} for regions in $(m_{K\pi},\delta m)$
dominated by signal (Fig.~\ref{fig:kpi-babar-time}(a))
and background (Fig.~\ref{fig:kpi-babar-time}(b)) events.

\begin{figure*}
  \center
  \psfig{figure=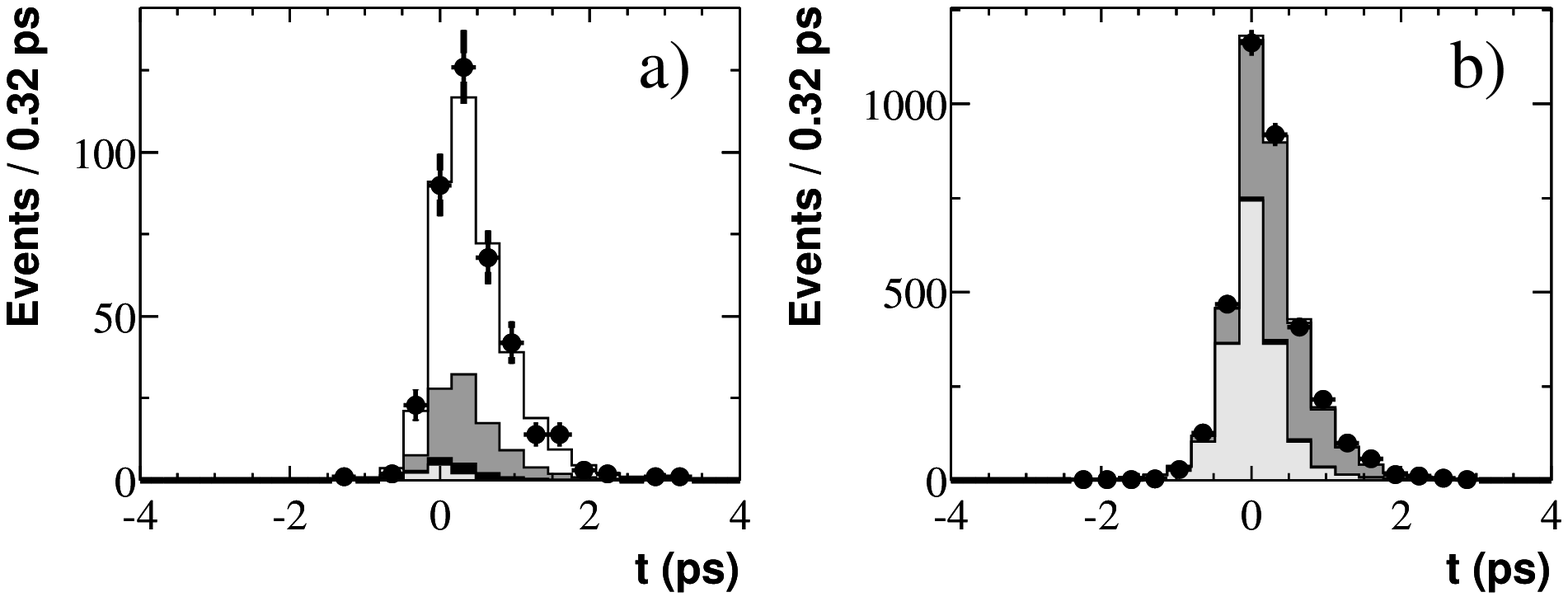,width=12.0truecm}
  \caption{BaBar $\dz\to K^+ \pi^-$ analysis:\protect\cite{kpi-babar}
	proper time distributions for (a) signal- and (b) background-dominated
	regions.
	Shown are the data (points), the projection of the fit (open histogram),
	and the fitted combinatorial (light), unassociated pion (dark),
	and double-misidentification background (black).}
  \label{fig:kpi-babar-time}
\end{figure*}

\begin{figure}
  \center
  \psfig{figure=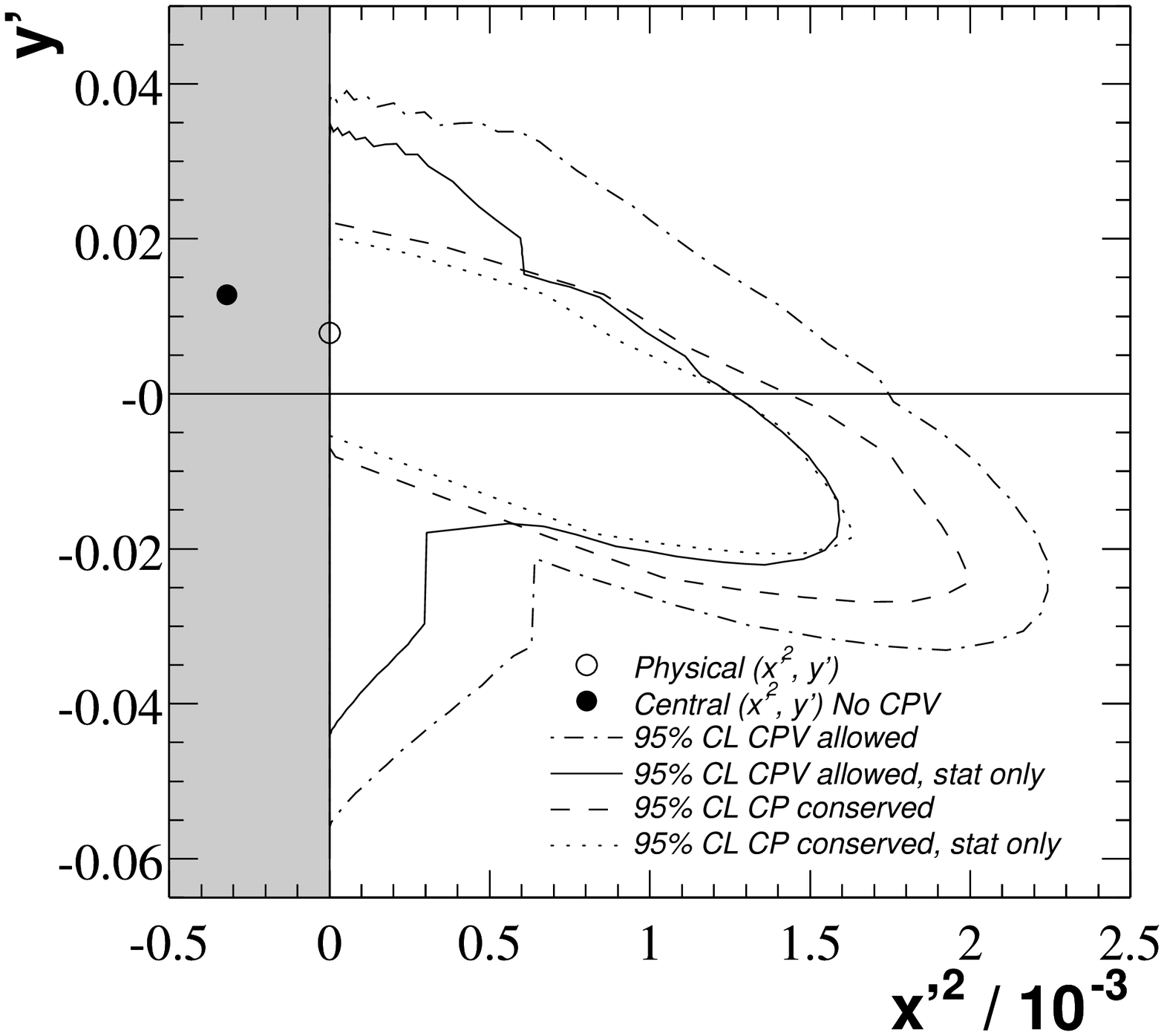,width=8.0truecm}
  \caption{BaBar $\dz\to K^+ \pi^-$ analysis:\protect\cite{kpi-babar}
	fit results and confidence intervals for the mixing parameters
	$(x^{\prime 2}, y^\prime)$.}
  \label{fig:kpi-babar-conf}
\end{figure}

Four overall fits allowing DCS decay only;
DCS decay and CP violation (treating \dz\ and \dzbar\ separately);
DCS decay and mixing, but no CP violation; and all three effects, are performed:
the results are shown in Table~\ref{tab:kpi-babar-results}.
No evidence for mixing or CP violation is found.
A slightly negative (and thus unphysical) value of $x^{\prime 2}$ is preferred by the fit:
this is taken into account in reporting the results.

\begin{tablehere}
  \caption{BaBar $\dz\to K^+ \pi^-$ analysis:\protect\cite{kpi-babar}
	parameters returned by the full fit to \dz, \dzbar, and combined samples.}
  \label{tab:kpi-babar-results}
  \renewcommand{\arraystretch}{1.4}
  \begin{tabular}{|lclll|}
  \hline
     Fit case & Para- & \multicolumn{3}{c|}{Fit result (${}/10^{-3}$)} \\
              & meter    &  \multicolumn{1}{c}{\hspace{3mm}\dz} 
 & \multicolumn{1}{c}{\hspace{3mm}\dzbar}
 & \multicolumn{1}{c|}{$\dz+\dzbar$}\\
  \hline
 \multirow{3}{1.5cm}{Mixing \\ allowed}
      & $R_{WS}^{(\pm)}$      & $\phantom{-0}3.9$   & $\phantom{-0}3.2$ & $\phantom{-}3.6$\\
      & ${x^{\prime(\pm)}}^2$ & $\phantom{0}$\relax$-0.79$  &
     $\phantom{0}$\relax$-0.17$ & $-0.32$ \\
      & $y^{\prime(\pm)}$      & $\phantom{-}17$    & $\phantom{-}12$ & $\phantom{-}13$ \\
    \hline
 No mixing
  & $R_{WS}^{(\pm)}$ & $\phantom{-0}3.9$ & $\phantom{-0}3.2$ & $\phantom{-}3.6$\\ 
    \hline
  \end{tabular}
\end{tablehere}

A rather careful procedure based on toy MC experiments is used to set frequentist
confidence intervals in the fitting parameters $x^{\prime 2}$ and $y'$: the results are shown
in Fig.~\ref{fig:kpi-babar-conf}. A remarkable feature of the analysis is that the allowed
region in $(x^{\prime 2},y')$ is comparable in size to that of CLEO, despite the larger and
cleaner dataset. Simulations by both BaBar\cite{kpi-ulrik} and Belle show that for a given experiment,
when the preferred value has $y' > 0$ (as in the BaBar analysis),
the allowed region becomes large compared to that when the data prefers $y' < 0$
(as in CLEO's case).\footnote{This may also be understood in qualitative terms
	using the proper time distribution for wrong-sign decays,
	\[ e^{-t} \left( R_D + \sqrt{R_D} y' t + \frac{x^{\prime 2} + y^{\prime 2}}{4} t^2
		  \right),
	\]
	where $R_D$ is the DCS decay rate. For $y' < 0 $ there is a partial cancellation
	between the interference ($\propto t e^{-t}$) and mixing ($\propto t^2 e^{-t}$)
	terms, and the fit becomes sensitive to small changes in both the $x^{\prime 2}$
	and $y'$ parameters. For $y' > 0$, no such cancellation takes place.}
It is thus almost meaningless to ``average'' the results of experiments whose
preferred values fall in different $y'$ regions.


\subsection{Mixing: Issues, Future Measurements}
\label{subsec:mixing-issues}

A combined $\dz \to K^+ \pi^-$ result will require a combined analysis:
perhaps there is a need for a joint working group, or at least
detailed consultation between experiments, as we enter the CLEO-c era. 
Comparison of $(x',y')$ and \ycp\ results
is more complicated still. The strong phase difference $\delta_{K\pi}$,
bequeathed us by the mischievous god of the color force, must first be measured:
a significant shift might occur due to final-state interactions (FSI),
shown to be significant in $D \to hh$ decays by isospin analyses of CLEO\cite{hh-cleo}
and FOCUS;\cite{hh-focus} the latter study provides evidence for inelastic FSI.

The default option is to wait for results from the CLEO-c run at the $\psi(3770)$,
as one of the analyses exploiting the coherent \dz\dzbar\ state
promises an error in $\cos\delta_{K\pi}$ of order $\pm 0.05$.\cite{cleo-c-gronau,cleo-c}
The other option is a complete measurement of the DCS $D \to K\pi$ decays, of which
only $\dz \to K^+\pi^-$ is currently known.
CLEO has recently placed a limit on $D^+ \to K^+ \pi^0$;\cite{hh-cleo} 
a measurement is presumably within the reach of the other $B$-factory experiments.
Measurement of $D^{+,0} \to \kz \pi^{+,0}$ rates relies on the
measurement of $D^{+,0} \to \kl \pi^{+,0}$ decays:
the asymmetry with the corresponding \ks\ mode is proportional to the interference between
decay amplitudes via \kz\ and \kzbar.
A method for this measurement has been demonstrated by Belle,
with a preliminary result for $\dz \to \kl \pi^0$.\cite{k0pi0-belle}

A promising new analysis method exploits the $\dz \to \ks \pi^+ \pi^-$ final state:
Cabibbo-favoured (e.g.\ $K^{\ast -}\pi^+$) and doubly-suppressed
($K^{\ast +}\pi^-$) decays interfere,
allowing measurement of the strong phase differences for the various resonant submodes;
a study of time-dependence then yields the mixing parameters $x$ and $y$.
This method has the advantages 
of superior scaling properties (the fit measures $x$ rather than $x^{\prime 2}$) and
sensitivity to the sign of the mass splitting,
in addition to the measurement of phases.
It is, however, unproven.
It has been championed by CLEO, who have measured the resonant substructure
of the decay;\cite{kspipi-cleo} a mixing study will presumably require
the large samples available at the other $B$-factories.
Those samples may also provide useful sensitivity from semileptonic decays,
which have fallen out of favor,
although there is an interesting unpublished
$\dz \to K^{(\ast)+} \ell^- \bar\nu_\ell$ analysis by CLEO.\cite{semilep-cleo}

Fits for CP violating effects in mixing are now routine,
and will increasingly become the main focus of study.
By the next Lepton Photon meeting, the other major development will be CLEO-c analyses
exploiting opposite-side tagging, geometric signal-to-background ratios,
and a coherent initial state leading to new mixing and CP violation observables.


\section{CPT and Lorentz Invariance Violation}
\label{sec:cpt}

Even more general analyses would allow for violation of CPT symmetry,
a manifest signal of new physics. While such studies have been carried out for
kaons and $B$ mesons, no CPT violation search had been performed in the charm
sector until the recent FOCUS analysis.\cite{cptv-focus}
Using \dstar-tagged $\dz\to K^-\pi^+$ decays, they searched for indirect CPT
violation parametrized by 
$\xi \equiv (\Lambda_{11} - \Lambda_{22}) / (\lambda_1 - \lambda_2)$,
where $\Lambda$ is the $2\times2$ effective Hamiltonian governing the time evolution
of neutral $D$ mesons, $\Lambda_{ii}$ are its diagonal elements, and $\lambda_j$ its
eigenvalues. The measured quantity is the time-dependent rate asymmetry
\[
  A_{\text{CPT}}(t)
  \equiv	\frac	{\Gamma(\dzbar \to K^+\pi^-) - \Gamma(\dz \to K^- \pi^+)}
			{\Gamma(\dzbar \to K^+\pi^-) + \Gamma(\dz \to K^- \pi^+)},
\]
which reduces to $\left( \myre(\xi) y - \myim(\xi) x \right) \Gamma t$
for $xt, yt \ll 1/\Gamma$;
$x$ and $y$ are the mixing parameters discussed above.
FOCUS finds $\left( \myre(\xi) y - \myim(\xi) x \right) = 0.0083 \pm 0.0065 \pm 0.0041$,
with a 95\% confidence interval
$-0.0068 < \left( \myre(\xi) y - \myim(\xi) x \right) < 0.0234$, 
consistent with zero. Their paper cites as an example the case where \dz\ and \dzbar\ 
mix with parameters $(x,y) = (0.0, 0.01)$: in this scenario, $0.68 < \myre \xi < 2.34$.

Within a formalism that allows for violation of Lorentz invariance,
$\xi$ may depend on the vector momentum of the studied particles,
and on siderial time; the relation is a function of parameters\cite{cptv-kostelecky}
$\Delta a_{0,X,Y,Z}$, where $(X,Y,Z)$ is a non-rotating
coordinate system. FOCUS fits for these quantities by measuring the CPT-violating
parameter $\xi$ in bins of siderial time: we can summarize the (complicated) result
by noting that the various $|\Delta a_\mu| < O\left(10^{-12}\right)\,\gev$ at 95\%
confidence for the case $(x,y,\delta_{K\pi}) = (0.01, 0.01, 15^\circ)$.
This is to be compared with limits of order $10^{-21}$ for the \kz-\kzbar\ system:
the difference reflects both the size of the available data samples
and the relative strength of mixing in the two systems.
The $\Delta a_\mu$ may in principle vary with flavor,
so this measurement is important for completeness,
even though the sensitivity does not compete with that for kaons.


\section{Rare and Forbidden Decays}
\label{sec:rare}

Moving to less exotic possibilities, flavor-changing neutral currents (FCNC)
have not yet been observed in the charm sector.
As with mixing, the SM parton-level loop contributions are subject to powerful
cancellations, and long-distance contributions dominate.
As a result, it can be difficult to calculate the SM expectation with certainty.
The best new physics signals are those decays with
extremely small predicted rates, or special features which allow SM and non-SM contributions
to be distinguished.

The decay $\dz \to \gamma \gamma$, with an expected branching fraction of order $10^{-8}$,
is thus a new physics signal (Sec.~\ref{subsec:rare-gammagamma});
but $\dz \to \mu^+ \mu^-$ (Sec.~\ref{subsec:rare-mumu}), whose predicted rate is  
five orders of magnitude smaller, is a more reliable one.\cite{hll-burdman}
In the case of decays $D \to h \ell^+ \ell^-$, the most robust new physics signal is not
the decay rate, but the dilepton mass spectrum, which exhibits marked differences
between SM and new-physics predictions (Sec.~\ref{subsec:rare-hll}). 
And in each case, variant modes violating lepton flavor or number conservation can be added
``for free'' to the analysis, in the spirit of a
lamp-post search: there is no uncertainty on Standard Model predictions of zero.
We consider each of these cases in turn,
before treating decays $\dz \to V\gamma$ in the final section 
(Sec.~\ref{subsec:rare-phigamma}).


\subsection{$\dz \to \gamma \gamma$ (CLEO)}
\label{subsec:rare-gammagamma}

\begin{figure}
  \center
  \psfig{figure=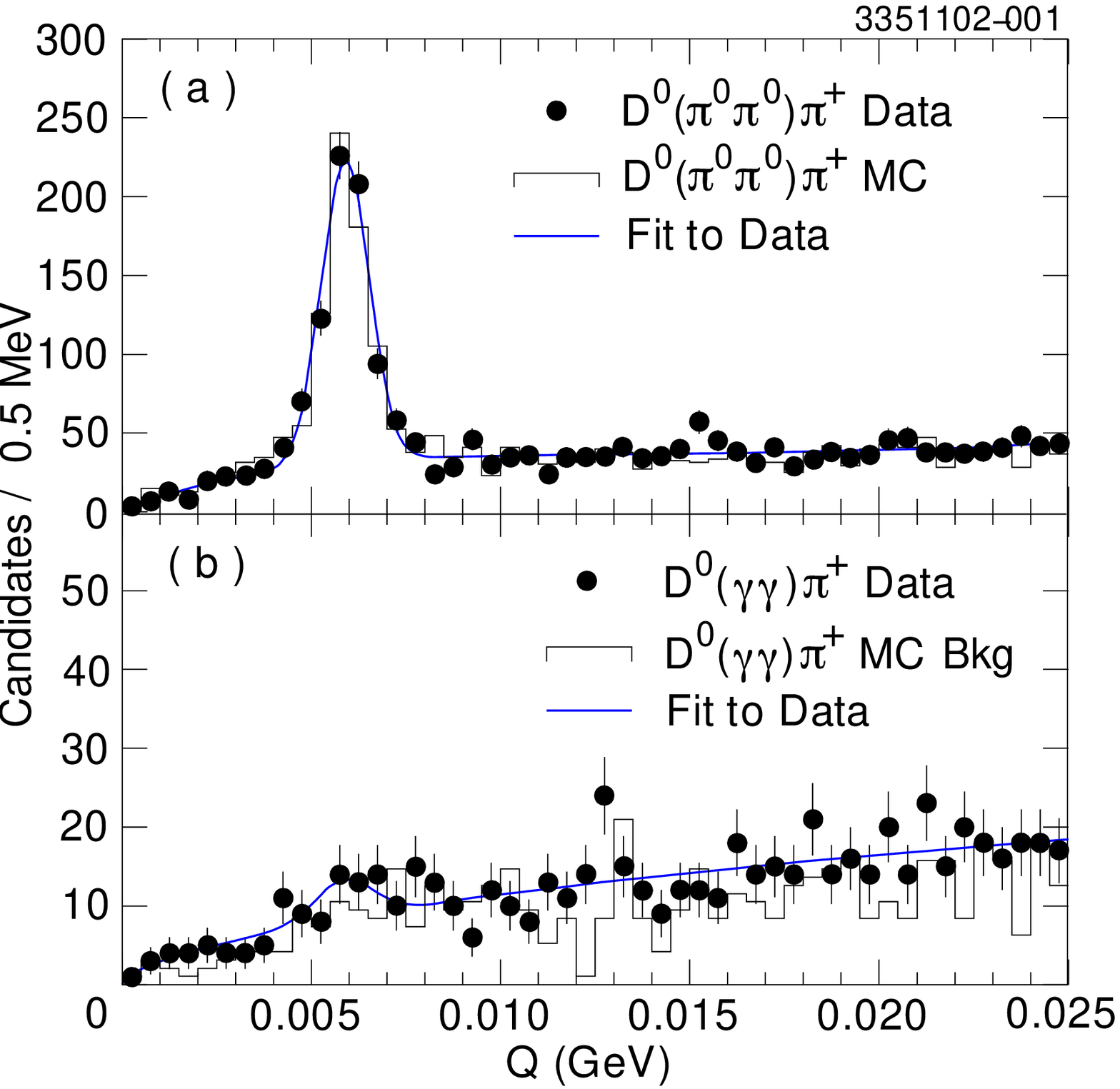,width=8.0truecm}
  \caption{CLEO $\dz\to \gamma\gamma$ analysis:\protect\cite{gamma-gamma-cleo}
  	$\dstarp-\dz$ mass-difference distributions
	for (a) $\dz\to \pi^0\pi^0$ and (b) $\dz\to \gamma\gamma$.}
  \label{fig:gamma-gamma-cleo}
\end{figure}

Decays to photons have very small contributions from parton-level
processes, but significant ones from the vector meson dominance (VMD) mechanism:
$\br(\dz \to \gamma \gamma)$ would be $\sim 3 \times 10^{-11}$ if only short-distance
mechanisms contributed, but is expected to be of order $10^{-8}$ due to long-distance SM
processes. This mode has not previously been studied. CLEO has recently used
$\dstarp \to \dz \pi^+$ events in $13.8\,\ifb$ of data to conduct a
$\dz \to \gamma\gamma$ search,
normalizing to the $\dz \to \pi^0 \pi^0$ mode.\cite{gamma-gamma-cleo}
Under standard cuts, augmented with a $\pi^0 \to \gamma\gamma$ veto on the photons forming
the $\dz \to \gamma\gamma$ candidates, they accumulate fairly clean event samples
and then fit the distribution of energy release $Q$ from the \dstar\ decay,
to keep differences between the $\pi^0 \pi^0$ and $\gamma\gamma$ modes to a minimum. 
The results are shown in Fig.~\ref{fig:gamma-gamma-cleo}: the agreement between
the data and the MC simulation is remarkable. 

A limit $\br(\dz \to \gamma\gamma) < 2.9 \times 10^{-5}$ is found
at the 90\% confidence level, some three orders of magnitude above the SM prediction. 
There is thus some room for a new physics signal if future experiments can improve 
on CLEO's sensitivity.


\subsection{$\dz \to \mu^+ \mu^-$ (CDF)}
\label{subsec:rare-mumu}

The expected rate for $\dz \to \mu^+ \mu^-$ is much smaller, $O(10^{-13})$,
whereas R-parity violating (RPV) Supersymmetry could lead to a branching fraction as high as
$3.5 \times 10^{-6}$, just smaller than the previous experimental bound.\cite{hll-burdman}
The dimuon decay is thus a straightforward new-physics search mode.
Using their upgraded detector and trigger system,
which allows them to select a charm decay sample,
CDF have conducted a search for this mode in the early Run II data.\cite{mumu-cdf}
With a fairly straightforward blind analysis,
using \dstar-tagged events and $\dz \to \pi^+ \pi^-$ decay as a normalization mode,
they observe no $\mu^+\mu^-$ events and set a limit
$\br(\dz \to \mu^+ \mu^-) < 2.4 \times 10^{-6}$ at 90\% confidence,
improving on the previous bound by a factor of two. 
The $O(1)$ event background estimate relies on interpolation from the sidebands---events
with true muon(s) dominate over the misidentification background---and will need to be
better understood to significantly improve the limit. 
Presumably this is achievable, and improvements to this channel will depend 
on the progress of Run II data-taking.


\subsection{$D \to h \ell \ell$ (FOCUS)}
\label{subsec:rare-hll}

The only analysis of a ``basket'' of rare decay modes in recent times is by FOCUS,
who have searched for decays
$D_{(s)}^+ \to h^\pm \mu \mu$, where $h^\pm = \pi^\pm,\, K^\pm$.\cite{hll-focus}
For definiteness we will take $D^+ \to \pi^+\mu^+\mu^-$ as an example.
The predicted SM rate for this decay is $O(10^{-6})$,
while (e.g.) the allowed R-parity violating contribution,\cite{hll-burdman}
$15 \times 10^{-6}$, saturates the previous experimental limit.\cite{hll-e791}
There would therefore seem to be potential for further restriction of RPV parameters,
but not for observation of a new physics signal.

\begin{figure}
  \center
  \psfig{figure=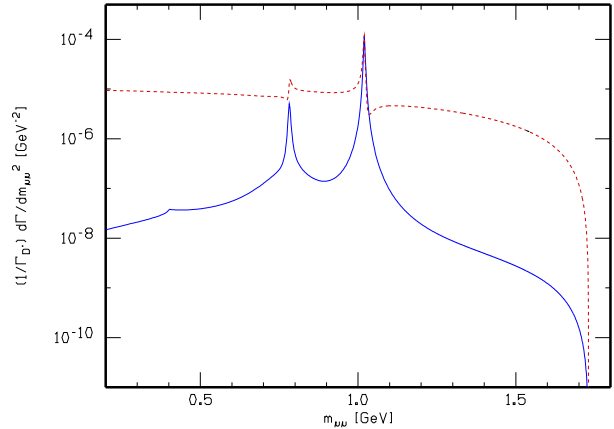,angle=90,width=8.0truecm}
  \caption{Predicted\protect\cite{hll-burdman} dimuon mass distributions
  	$M(\mu^+\mu^-)$ for $D^+ \to \pi^+ \mu^+ \mu^-$.
	The solid line shows the sum of the short- and long-distance contributions
	in the SM; the dotted line, the R-parity violating contribution from SUSY
	at the level allowed prior to the FOCUS measurement---see the text.}
  \label{fig:hll-theory-pi+mu+mu-}
\end{figure}

The SM contribution, however, is dominated by the path $D^+ \to \pi^+ V \to \pi^+ \mu^+ \mu^-$, where the $V$ are vector mesons. 
The predicted dimuon mass spectrum (Fig.~\ref{fig:hll-theory-pi+mu+mu-})
thus shows pronounced peaks at the $\rho$ and $\phi$ masses, which dominate the SM rate.
By contrast the spectrum for RPV is relatively flat,
so that a new physics contribution comparable to or even below the SM contribution could
be resolved  by comparing the  $M(\mu^+\mu^-)$ distribution of observed events with the
various predictions.

\begin{figurehere}
  \center
  \psfig{figure=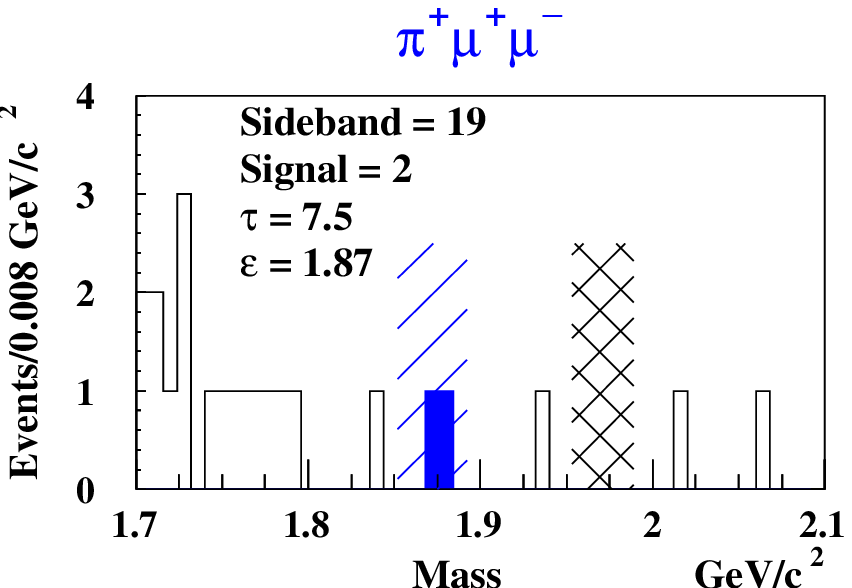,width=8.0truecm}
  \caption{FOCUS $D^+ \to \pi^+ \mu^+ \mu^-$ analysis:\protect\cite{hll-focus}
  	the histogram shows selected events; the filled entry is in the signal region.
	The signal region (hatched) and the region used for sideband subtraction
	(cross-hatched) are also shown.}
  \label{fig:hll-focus-pi+mu+mu-}
\end{figurehere}

FOCUS is beginning to probe this region.
A histogram of selected events for the $D^+ \to \pi^+\mu^+\mu^-$ analysis is shown in
Fig.~\ref{fig:hll-focus-pi+mu+mu-}, and results for all the modes are listed in 
Table~\ref{tab:hll-focus}. The analysis proceeds using standard FOCUS detached vertex,
hadron- and muon-identification cuts;
cut values are selected and background estimates calculated using a very careful
``dual bootstrap'' method, to minimise possible selection biasses.
No evidence is seen for any of the decay modes;
existing limits are everywhere improved, in some cases by an order of magnitude.
For the $D^+ \to \pi^+\mu^+\mu^-$ and $D_s^+ \to \pi^+\mu^+\mu^-$ modes, the sensitivity
is approaching the SM prediction.

\begin{tablehere}
  \caption{FOCUS $D \to h \ell \ell$ analysis:\protect\cite{hll-focus}
  	measured limit on the branching fraction, SM prediction,\protect\cite{hll-fajfer}
	previous best limit,\protect\cite{hll-e791,hll-e687}
	and expected CLEO-c sensitivity\protect\cite{cleo-c}
	for each mode.
	($D_s^+$ sensitivities are scaled from those of $D^+$ and are not
	official CLEO-c numbers.)
	All entries are ($/ 10^{-6}$).}
  \label{tab:hll-focus}
  \renewcommand{\arraystretch}{1.2}
  \begin{tabular}{|lcccc|}
    \hline
    Mode			& FOCUS	& SM	& Prev.		& CLEO-c	\\
    \hline
    $D^+  \to K^+\mu^-\mu^+$	& $9.2$ &$0.007$& $44$		& $1.5$		\\
    $D^+  \to K^-\mu^+\mu^+$	& $13$  &   -	& $120$		& $1.5$		\\
    $D^+  \to \pi^+\mu^-\mu^+$	& $8.8$ & $1.0$	& $15$		& $1.5$		\\
    $D^+  \to \pi^-\mu^+\mu^+$	& $4.8$ &   -	& $17$		& $1.5$		\\
    $D_s^+\to K^+\mu^-\mu^+  $	& $36$  &$0.043$& $140$		& $15$		\\
    $D_s^+\to K^-\mu^+\mu^+$	& $13$  &   -	& $180$		& $15$		\\
    $D_s^+\to \pi^+\mu^-\mu^+$	& $26$  & $6.1$	& $140$		& $15$		\\
    $D_s^+\to \pi^-\mu^+\mu^+$	& $29$  &   -	& $82$		& $15$		\\
    \hline
\end{tabular}
\end{tablehere}

Further progress is expected at CLEO-c, whose sensitivities\cite{cleo-c} are also shown:
an improvement by a factor $3 \sim 8$ is foreseen for the $D^+$ modes,
reaching the SM expectation in the case of $D^+ \to \pi^+\mu^+\mu^-$
and therefore restricting further the RPV contribution.
Any signal in the lepton-number violating modes
$D_{(s)}^+ \to h^- \mu^+ \mu^+$, albeit unexpected, would of course be an observation
of new physics.


\subsection{$\dz \to \phi \gamma,\, \phi\pi^0,\, \phi\eta$ (Belle)}
\label{subsec:rare-phigamma}

Finally we turn to radiative decays $\dz \to V \gamma$,
another vector meson dominance process in the Standard Model. 
The Belle collaboration has conducted a search for $\dz \to \phi \gamma$,
exploiting double kaon identification in $\phi \to K^+ K^-$ to suppress
backgrounds.\cite{phigamma-belle} 
Theoretical estimates for this mode,\cite{phigamma-burdman,phigamma-fajfer}
dominated by $\dz \to V V' \to V \gamma$ (where the $V^{(\prime)}$
are vector mesons), are in the range $(0.04 \sim 3.4) \times 10^{-5}$,
well below the previous limit of $1.9 \times 10^{-4}$ but partially overlapping
the sensitivity at the $B$-factories.

\begin{figure}
  \center
  \psfig{figure=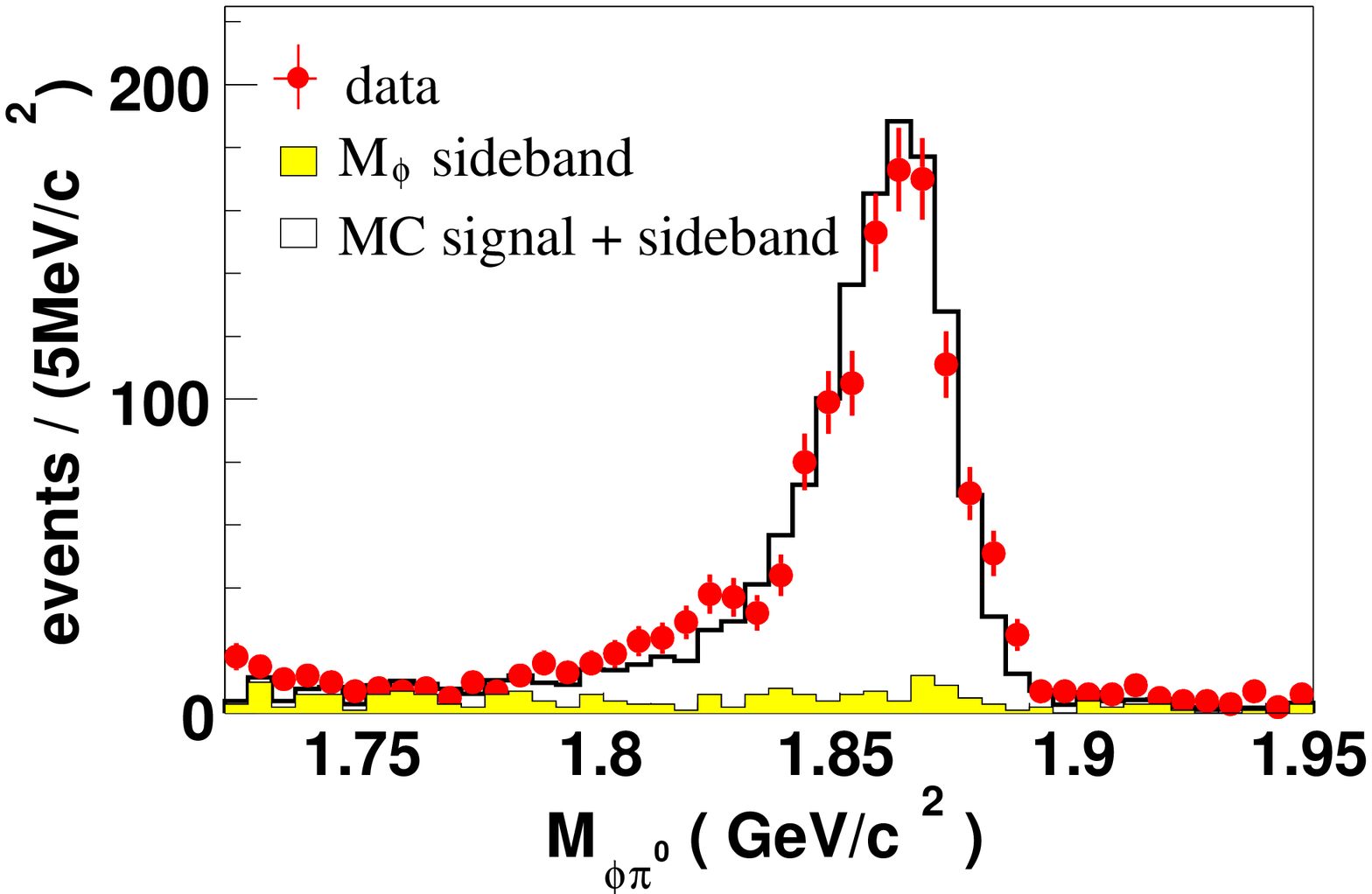,width=8.0truecm}
  \caption{Belle $\dz\to \phi\gamma,\, \phi\pi^0,\, \phi\eta$ analysis:\protect\cite{phigamma-belle}
  	$\phi\pi^0$ mass distribution for data (points) and MC (histogram);
	the $\phi$-mass sideband is also shown (shaded).}
  \label{fig:phipi0-mass}
\end{figure}

\dstar-tagging and cuts on the \dstar\ momentum and $\gamma$ energy are used to 
suppress the various combinatorial backgrounds.
The dominant remaining background is due to the Cabibbo-
and color-suppressed decays $\dz \to \phi \pi^0$ and $\phi \eta$,
which have not previously been observed. 
With analagous cuts to select $\dz \to \phi \pi^0$ Belle sees a very clear 
signal in $M(\phi\pi^0)$ (Fig.~\ref{fig:phipi0-mass}) and the expected distribution
of the helicity angle of the $\phi$ meson (not shown); the $\phi$ is polarized in
the \dz\ decay. A smaller but still clear signal of $31 \pm 9.8$ events is seen for
$\dz \to \phi \eta$, where a veto is imposed on photons for the $\eta \to \gamma \gamma$
candidate consistent with belonging to a $\pi^0 \to \gamma \gamma$ decay.

\begin{figurehere}
  \center
  \psfig{figure=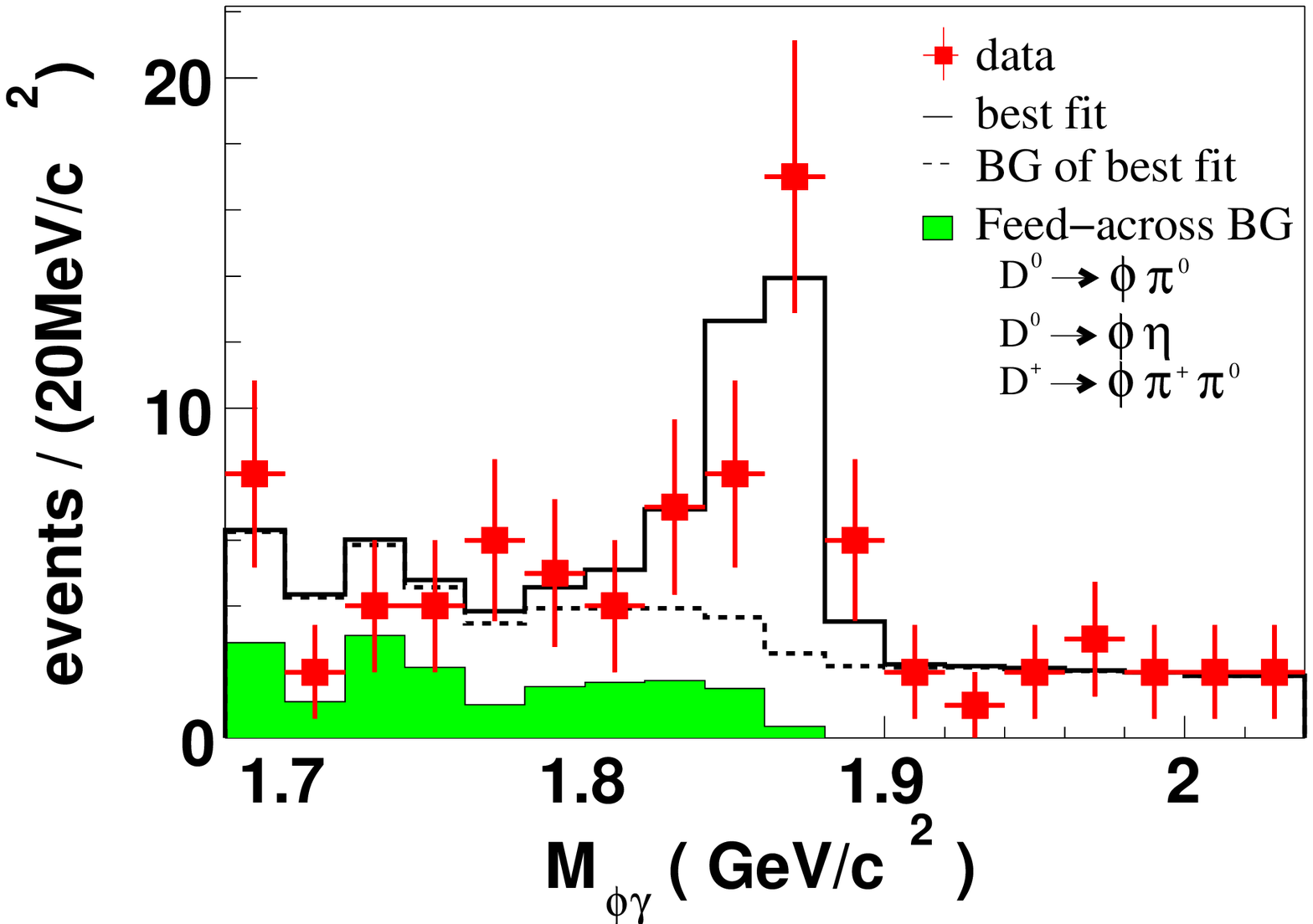,width=8.0truecm}
  \caption{Belle $\dz\to \phi\gamma,\, \phi\pi^0,\, \phi\eta$ analysis:\protect\cite{phigamma-belle}
  	$\phi\gamma$ mass distribution for data (points),
	the ML fit (open histogram),
	the background component of the fit (dashed), and
	the sum of $\dz\to \phi\pi^0,\, \phi\eta$ 
	and $D^+ \to \phi\pi^+\pi^0$ backgrounds (shaded).}
  \label{fig:phigamma-mass}
\end{figurehere}

The contribution of these decays to the $\phi\gamma$ spectrum can then be reliably
estimated; it is suppressed by a helicity angle cut $|\cos\thel| < 0.4$,
favoring the transversely polarized $\phi$ of $\dz\to\phi\gamma$ over the longitudinally
polarized $\phi$ of the $\phi\pi^0$ and $\phi\eta$ modes.
The resulting $\phi\gamma$ invariant mass spectrum shows a clear 
$\dz\to \phi\gamma$ signal of $27.6^{+7.4}_{-6.5} {}^{+0.5}_{-1.0}$ events
(Fig.~\ref{fig:phigamma-mass}),
and is well-described by a linear combinatorial background,
the expected $\phi\pi^0$ and $\phi\eta$ contribution, plus the signal.
The helicity angle distribution (Fig.~\ref{fig:phigamma-helicity})
is likewise consistent with expectations.

%
%

The final results of the analysis are
\begin{align*}
  \br(\dz \to \phi \pi^0)	& = (8.01 \pm 0.26 \pm 0.46) \times 10^{-4}	\\
  \br(\dz \to \phi \eta)	& = (1.48 \pm 0.47 \pm 0.09) \times 10^{-4}	\\
  \br(\dz \to \phi \gamma)	& = \left( 2.60^{+0.70}_{-0.61} {}^{+0.15}_{-0.17} \right) \times 10^{-5};
\end{align*}
the $\phi\gamma$ mode is the first radiative decay, and the first FCNC decay,
observed in the $D$ meson system. The measured branching fraction is at the upper
end of the VMD predictions---consistent with Standard Model expectations---and is in
no sense a new physics measurement, but it provides the first experimental reference point
for predictions of other FCNC decays.

\begin{figure}
  \center
  \psfig{figure=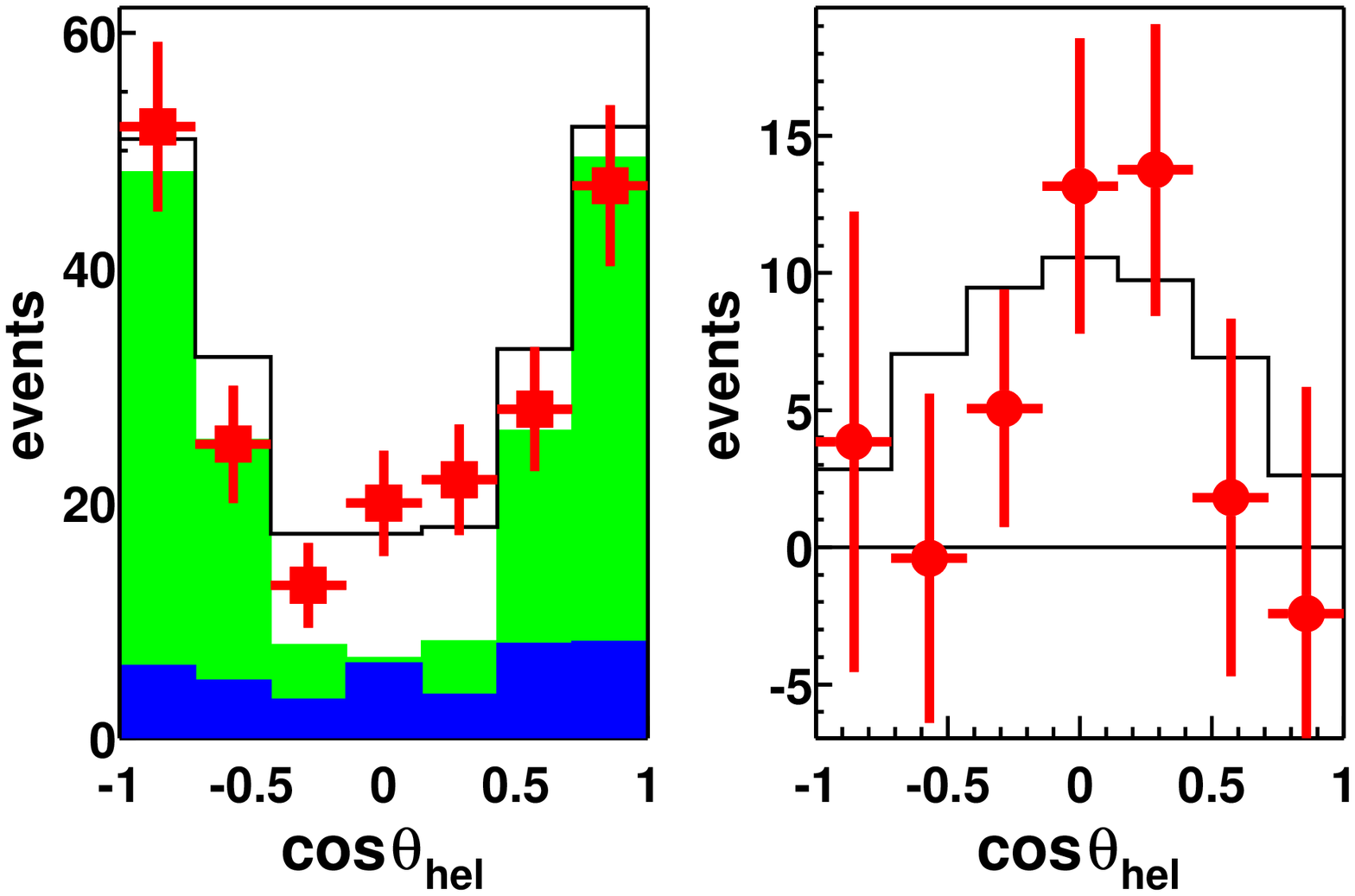,width=8.0truecm}
  \caption{Belle $\dz\to \phi\gamma,\, \phi\pi^0,\, \phi\eta$ analysis:\protect\cite{phigamma-belle}
  	cross-check of the $\phi\gamma$ result using the $\cos\thel$ distribution.
	Left plot: data (points), and MC predictions for the total (open histogram),
	total background (light), and non-$\phi\pi^0$ background (dark).
	Right plot: background-subtracted distribution for the data (points)
	and the MC prediction (histogram).}
  \label{fig:phigamma-helicity}
\end{figure}


\section{Summary and Prospect}
\label{sec:summary}

While all observations are still consistent with Standard Model expectations,
there has been significant progress in charm analyses sensitive to new physics
effects. The $D$-mixing measurement using $\dz \to K^+\pi^-$ is now mature,
with a major result released by BaBar, and a Belle analysis in the pipeline;
the $B$-factories will (presumably) exhaust this difficult technique.
The sensitivity to \ycp\ continues to improve, and a 1\% measurement is within
reach of the current facilities during their projected life.
Fitting for CP-violating effects has become standard, and as a robust test for
new physics---and a technique still dominated by statistical errors---will
become increasingly important in the field.
And measurements of the new quantities observable at CLEO-c,
and the promising but yet-untried $\dz \to \ks \pi^+ \pi^-$ mixing analysis,
are eagerly awaited.

The first flavor-changing neutral current decay in the charm sector,
$\dz \to \phi\gamma$, has finally been seen, and is consistent with Standard
Model expectations due to vector meson dominance.
While not the most exciting possible channel---$\gamma\gamma$ or $\mu^+\mu^-$
would have been respectively a shock, and a total revolution in
the field---the $\phi\gamma$ observation provides an experimentally-measured
point where none previously existed, and will presumably allow more precise
SM predictions for other FCNC decays in the future.
In the search for other rare and exotic processes, there are continuing 
improvements in the range of channels studied
(with the first $\dz \to \gamma \gamma$ limit announced,
and the first test of CPT/Lorentz invariance conducted)
and the reach of existing analyses (with the $D \to h \ell \ell$ results
from FOCUS). As with mixing, there will be significant contributions to these
searches from CLEO-c in the next two years, and we all---including even the
charm coordinators at BaBar, Belle, CDF, and FOCUS---are looking forward to 
new results in this new era.

\section*{Acknowledgments}
I would like to thank the Lepton Photon 2003 Symposium organisers for the
invitation to present this review;
Lin Zhang, Marc Buehler, and Harry Cheung for their technical assistance
with the presentation itself and this writeup;
and my colleagues in the Belle collaboration,
in particular the charm studies group,
for providing such a stimulating working environment.

\balance

\clearpage
\twocolumn[
\section*{DISCUSSION}
]

\begin{description}
  \item[Hal Evans] (Columbia):
	What is the average \ycp\ measurement?
	Or is there a reason not to calculate it?
  \item[Bruce Yabsley:]
	The average would be dominated by preliminary measurements from
	BaBar and Belle. Further, results appear to be ``clumping'' based
	on measurement technique indicating the possibility of systematic
	problems. That being said, the speaker's average is
	$\langle \ycp \rangle = (0.9 \pm 0.4)\%$
\end{description}

\subsection*{Further Discussion, Added since LP2003:}

It's appropriate to return to this question in more detail
for the written version of the talk.
I reserved the \ycp\ average for the Discussion---assuming, correctly,
that someone would bring it up---partly for lack of time
and partly for lack of a clear idea of how to treat it:
any average is dominated by a preliminary number from Belle\cite{ycp-belle-tag}
and results from BaBar which, though ``final'', had not been published
at the time of the symposium. Since then, BaBar's paper has appeared
in \emph{PRL},\cite{ycp-babar} so my reservations are now diminished.
The situation, however, is still unclear.

\begin{tablehere}
  \caption{Expanded summary of \ycp\ results.}
  \label{tab:ycp-summary-extended}
  \begin{center}
    \renewcommand{\arraystretch}{1.2}
    \begin{tabular}{|lll|}
	\hline
	\textbf{Technique}	&  \textbf{Experiment}	& \multicolumn{1}{c|}{$\phantom{--}\mathbf{\ycp\;(\%)}$}\\
	\hline
	\multirow{3}{1.8cm}{\textbf{\noindent
	  Fixed target}}	&  E791\cite{ycp-e791}			& $\phantom{-}0.8 \pm 2.9 \pm 1.0$	\\
				&  FOCUS\cite{ycp-focus}		& $\phantom{-}3.4 \pm 1.4 \pm 0.7$	\\
				&  \textbf{\emph{My average:}}		& $\phantom{-}\mathbf{
										      2.9 \pm 1.4}$		\\
	\hline
	\multirow{4}{1.8cm}{\textbf{\boldmath
	  $e^+e^-$, untagged}}	& Belle\cite{ycp-belle-untag}		& $          -0.5 \pm 1.0 \pm 0.8$	\\
	  			&  CLEO\cite{ycp-cleo}			& $          -1.2 \pm 2.5 \pm 1.4$	\\
				&  BaBar\cite{ycp-babar}		& $\phantom{-}0.2 \pm 0.5^{+0.5}_{-0.4}$\\
				&  \textbf{\emph{My average:}}		& $\phantom{-}\mathbf{
										      0.0 \pm 0.6}$		\\
	\hline
	\multirow{4}{1.8cm}{\textbf{\boldmath
	  $e^+e^-$, \dstar-tagged}}
				& BaBar,\cite{ycp-babar} $K^+K^-$	& $\phantom{-}1.5 \pm 0.8 \pm 0.5$	\\
				& BaBar,\cite{ycp-babar} $\pi^+\pi^-$	& $\phantom{-}1.7 \pm 1.2^{+1.2}_{-0.6}$\\
				& Belle,\cite{ycp-belle-tag} $K^+K^-$	& $\phantom{-}1.2 \pm 0.7 \pm 0.4$	\\
				& \textbf{\emph{My average:}}		& $\phantom{-}\mathbf{
										      1.4 \pm 0.6}$		\\
	\hline
	\multicolumn{2}{|l}{\textbf{\emph{\noindent
	  Speaker's grand average:}}}					& $\phantom{-}\mathbf{
	  									      0.9 \pm 0.4}$		\\
	\hline
  \end{tabular}
  \end{center}
\end{tablehere}

The results from the various experiments are shown
in Table~\ref{tab:ycp-summary-extended}, sorted by the type of experiment and
the method used: fixed target (E791\cite{ycp-e791} and FOCUS\cite{ycp-focus}),
$e^+e^-$ with inclusive \dz\ samples (Belle,\cite{ycp-belle-untag} CLEO,\cite{ycp-cleo}
and BaBar\cite{ycp-babar}), and $e^+e^-$ with \dstar-tagged samples
(BaBar\cite{ycp-babar} and Belle\cite{ycp-belle-tag}).
The individual results within the BaBar analysis have been listed separately for this
purpose. There is a clear clustering of the results according to technique:
fixed target measures high ($\langle\ycp\rangle = 2.9\%$),
$e^+e^-$ measures null ($\langle\ycp\rangle = 0.0\%$),
and the \dstar-tagged analyses measure an intermediate value ($\langle\ycp\rangle = 1.4\%$).
These last results are completely dominant, as can be seen
in Fig.~\ref{fig:ycp-ideogram}, where the data are shown in the form of 
the ``ideograms'' used by the PDG.\cite{pdg}
\begin{figure}[h!]
  \center
  \psfig{figure=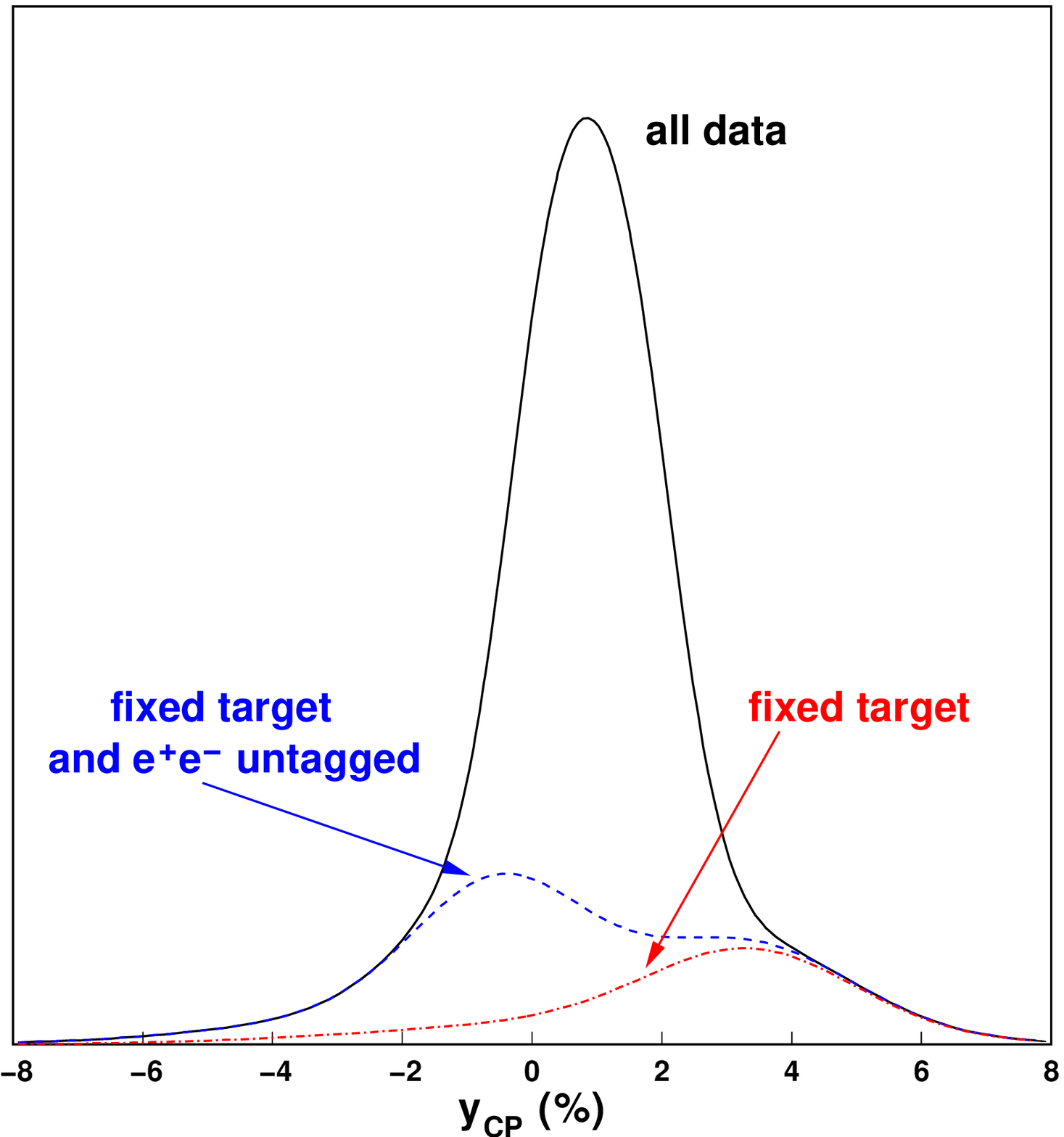,width=5.8truecm}
  \vspace{-2mm}
  \caption{PDG-style ``ideogram'' of the \ycp\ data
  	in Table~\protect\ref{tab:ycp-summary-extended}:
	each measurement (with mean $x_i$ and total error $\pm \delta x_i$)
	is represented by a gaussian with central value $x_i$, error $\delta x_i$,
	and area proportional to $1/\delta x_i$. The sum of all curves (solid),
	the sum of fixed-target and $e^+ e^-$ untagged data (dashed),
	and the fixed-target data alone (dot-dashed) are shown.}
  \label{fig:ycp-ideogram}
\end{figure}
\par Statistically speaking, the data are consistent, and the average
is now $2\sigma$ away from zero. But it seems to me a bit worrying that the
different techniques don't actively corroborate each other.
Put another way, having first become excited about the (positive) FOCUS result, 
and then rushed to discount it in the light of (negative) $e^+e^-$ results,
I think we should be slow to interpret \dstar-tagged results
which split the difference.

This question will become urgent after the next round of data-taking at
the $B$-factories, if the central value stays at $\ycp \gtrsim 1\%$ as the 
total error shrinks. One convincing cross-check would
be to analyse \dz\ decays to CP-odd final states such as $\ks (\rho^0, \omega, \phi)$
and even $\ks \eta^{(\prime)}$
at the $B$-factories: for the same \ycp\ value, the shift in the lifetime has
the opposite sign. I know of no such plans at this stage.

\end{document}